\documentclass[12pt]{article}

\def\slash#1{\setbox0=\hbox{$#1$}#1\hskip-\wd0\hbox to\wd0{\hss\sl/\/\hss}}
\usepackage{epsf, cite}
\usepackage{epsfig}                  
\usepackage[all]{xy}                 
\usepackage{amsfonts}
\usepackage{latexsym}

\usepackage{epsf, cite}

\makeatletter
\renewcommand\section{\@startsection {section}{1}{\z@}%
                                   {-3.5ex \@plus -1ex \@minus -.2ex}
                                   {2.3ex \@plus.2ex}%
                                   {\normalfont\large\bfseries}}
\renewcommand\subsection{\@startsection{subsection}{2}{\z@}%
                                     {-3.25ex\@plus -1ex \@minus -.2ex}%
                                     {1.5ex \@plus .2ex}%
                                     {\normalfont\bfseries}}
\makeatother

\def\lbldef#1#2{\expandafter\gdef\csname #1\endcsname {#2}}

\def\href#1#2{#2}

\newcommand{\beq}{\begin{equation}}
\newcommand{\eeq}{\end{equation}}
\def\wv{worldvolume }

\def\bea{\begin{eqnarray}}
\def\eea{\end{eqnarray}}
\newcommand{\pd}[2]{\frac{\partial{#1}}{\partial{#2}}}

\def\gs{\,\raise.15ex\hbox{/}\mkern-11.5mu G} 
\def\cs{\,\raise.15ex\hbox{/}\mkern-11.5mu C} 

\def\lt{\lfloor}
\def\rt{\rfloor}
\newcommand{\tb}[3]{[X^{#1},X^{#2},X^{#3}]}
\newcommand{\gtb}[3]{[H^*,X^{#1},X^{#2},X^{#3}]}
\newcommand{\fbr}[5]{\lt X^{#1},X^{#2},\tb{#3}{#4}{#5}\rt}

\def\Prp{ { {\cal{P}}_{\cal{R^+}} }  } 
\def\Prm{  {\cal{P}}_{ {\cal{R^-}} }  }
\def\cRp{ { { \cal{R}}^+}  } 
\def\cRm{ { {\cal{R}}^-}  }
\def\cR{ {\cal{R}}  }

\providecommand{\openone}{\leavevmode\hbox{\small1\kern-3.8pt\normalsize1}}

\begin{document}

\setcounter{page}{1}



\begin{titlepage}

\begin{center}

\hfill QMUL-PH-05-03\\ 
\hfill DAMTP-2005-36

\vskip 2 cm
{\Large \bf 
Five-brane Calibrations and Fuzzy Funnels}\\

\vskip 1.25 cm {David S. Berman\footnote{email: D.S.Berman@qmul.ac.uk}}
\\
{\vskip 0.5cm
Queen Mary College, University of London,\\
Department of Physics,\\
Mile End Road,\\
London, E1 4NS, England.}\\
{\vskip 0.5cm
and}

\vskip 0.5 cm {Neil B. Copland\footnote{email: N.B.Copland@damtp.cam.ac.uk}}
\\
{\vskip 0.5cm
Department of Applied Mathematics and Theoretical Physics,\\
Centre for Mathematical Sciences,\\University of Cambridge,\\
Wilberforce Road,\\
Cambridge CB3 0WA,\\ England.}\\

\end{center}   

\vskip 2 cm

\begin{abstract}  
\baselineskip=18pt

We present a generalisation of the Basu-Harvey equation that 
describes membranes ending on intersecting five-brane configurations 
corresponding to various calibrated geometries.

\end{abstract}

\end{titlepage}

\pagestyle{plain}

\baselineskip=19pt

\section{Introduction}

There are many different but physically equivalent descriptions of how
a D1 brane may end on a D3 brane. From the point of view of the D3
brane the configuration is described by a monopole on its world
volume. From the point of view of the D1 brane the configuration is
described by the D1 opening up into a D3 brane where the extra two
dimensions form a fuzzy two sphere whose radius diverges at the origin
of the three-brane. These different view points are the stringy
realisation of the Nahm transformation. The BPS equation obeyed by the
D1 brane is Nahm's equation. These differing perspectives on the D1, D3
system  have been explored in a variety of papers \cite{Callan,Gibbons,Brecher,Myers,Cook,Constable} where the
fluctuation  properties, the profile and the coupling to RR fields are
examined and shown to match where the different approximations schemes
are both valid.

The M-theory equivalent of this system is that of the membrane ending
on the  M five-brane. There are well known problems though of
describing the theory of coincident branes in M-theory, both for the
membrane and the five-brane. It is believed that matrix valued fields
will not be the appropriate degrees of freedom as they are for
D-branes since entropy considerations imply a different number of
degrees of freedom than one would expect from matrix valued fields.

What is known is the five-brane equivalent of the BIon solution. This
is the self-dual string solution of Howe, Lambert and West
\cite{HLW} whose properties have recently been investigated in \cite{Berman}. 
The absence of a non-Abelian membrane theory however meant
that the equivalent of Nahm's equation for the self-dual string was
missing. Recently, Basu and Harvey \cite{Basu:2004ed}, made an ansatz for 
such an
equation. Their goal being to produce a similar fuzzy funnel
description of the membrane opening up into the five-brane. From their
generalised Nahm equation they went on to infer (essentially through
an inverse Bogomol'nyi argument) a non-Abelian membrane action with
sextic interaction. Again, the profile and the fluctuations were shown
to agree with the self-dual string description.

Given the somewhat ad-hoc way in which the Basu-Harvey equation has
been determined it would be good to see if one could find other tests
for its validity. The goal of this paper will be to show that the
Basu-Harvey equation with a natural generalisation can describe not
only the membrane ending on a five-brane but also the membrane ending
on various intersecting five-brane configurations. These five-brane
configurations will correspond to calibrated cycles. 
Essentially this is the M-theory generalisation of
\cite{Constable:2002yn} where the  Nahm equation was generalised to
describe the D1 ending on the configurations of D3 branes that
correspond to three-branes wrapped on calibrated cycles.

We proceed by first describing Nahm's equation and the generalisation
that  leads to the description of a D1 ending on intersecting D3-brane
configurations. As an aside we also demonstrate that the solutions
previously derived using the linearised approximation in
\cite{Constable:2002yn} are actually solutions of the full non-linear
theory. This is undoubtedly a consequence of the BPS nature of these
solutions. Then we discuss the M-theory generalisation of Nahm's
equation as introduced by Basu and Harvey. Finally we describe a
generalisation of the Basu-Harvey equation and its solutions that
correspond to membranes ending on five-branes wrapped on calibrated
cycles.

\section{Nahm Type Equations}

The Nahm equation \cite{Nahm} is given by \beq\label{Nahm}
\pd{\Phi^i}{x^9}=\pm \frac{i}{2}\epsilon_{ijk}[\Phi^j,\Phi^k].  \eeq
This equation is derived in string theory by simply examining  the
$\frac{1}{2}$ BPS equation for the D1 brane \cite{Diaconescu,Tsimpis}.
Its solutions correspond to D1 branes opening up into a fuzzy funnel
to form  a D3 brane. Note, the boundary conditions are taken such that
$\Phi^i(x^9)$ is  defined over the semi-infinite line as opposed to a
finite interval which is  the usual case. This corresponds to
infinite mass monopoles which have the  interpretation of infinite D1
strings ending on the brane. Explicitly, the  solutions are:
\beq\label{Nahmsol} 
\Phi^i= \pm  \frac{1}{2(\sigma-\sigma_0)} \alpha^i \,\, .  
\eeq
Where $\alpha^i$ obey the $su(2)$ algebra,
$[\alpha^i,\alpha^j]=2i\epsilon^{ijk} \alpha^k$. The sign choice is
related to whether the solution is BPS or anti-BPS, in what follows a
particular sign will be chosen though one should keep in mind that
one can choose the opposite sign and the resulting solutions will
simply be the anti-BPS equivalent.  Nahm's equation was generalised in
\cite{Constable:2002yn} by considering not just  the $\frac{1}{2}$ BPS
equation of the D1 brane but instead by looking at  the BPS equation
that arises from preserving a lower number of supersymmetries. This
produced the following generalised Nahm equation: 
\beq\label{ModNahm}
\pd{\Phi^i}{x^9}=\pm \frac{i}{2}c_{ijk}[\Phi^j,\Phi^k], \eeq where
$c_{ijk}$ is some totally anti-symmetric constant tensor with 
$i,j,k =1,\ldots,\rm{d}$. 
Along with this Nahm type equation there is also a set of algebraic equations 
that arise from the BPS conditions. This generalised Nahm equation along with 
the algebraic equations together imply the equations of motion. (We will not 
give the algebraic equations here since they are dependent on the details of 
the preserved supersymmetry, see \cite{Constable:2002yn} for details).

Spinors that obey the supersymmetry projection conditions may be
used to write down a calibration form,  $c=c_{ijk} dx^i \wedge dx^j
\wedge dx^k$ using $c_{ijk}= {\bar{\epsilon}} \Gamma_{ijk}
\epsilon$. The  components of this form  $c_{ijk}$ are then what
appear in the modified Nahm equation given above. (It should be
stated that this modified Nahm equation with a specific $c$ has appeared 
in the pre D-brane literature, for example \cite{Fairlie et al}). The 
solutions to this equation then correspond to D1 branes ending on a three-brane
that wraps the calibrated cycle or equivalently a set of intersecting
three-branes. The description of the relation between calibrated
cycles and intersecting brane configurations in string theory was
described in \cite{Gibbons:1998hm,Jerome}.

The analysis performed in \cite{Constable:2002yn} examined the
linearised D1 brane action. Here we examine the full non-Abelian 
Born-Infeld action for the D1, given by
\beq
S=-T_1\int d^2\sigma
STr\sqrt{-det(\eta_{ab}+\lambda^2\partial_a\Phi^i Q_{ij}^{-1}\partial_b\Phi^j)det(Q^{ij})},
\eeq
where  $Q^{ij}=\delta^{ij}+i\lambda[\Phi^i,\Phi^j]$ and 
$\lambda= 2 \pi l_s^2 $. STr denotes
the symmetrised trace prescription \cite{Tseytlin}. In 2-dimensions
the gauge field carries no propagating degrees of freedom and may be
completely gauged away, which is why only partial derivatives appear
in the above action. (It is known that
there is some possible ambiguity in the non-Abelian Born-Infeld theory
since the derivative approximation is not valid in a non-abelian
theory yet this action has been shown to possess many of the right
properties, see for example \cite{Myers,Brecher}).

We now wish to write down the energy so as to obtain a Bogomol'nyi style 
argument. If we restrict ourselves to a static solution with three non-zero
scalars, which only depend on $\Phi^9$ we can expand the determinant
to give an expression for the energy \beq E=T_1\int d\sigma STr\sqrt{I+\lambda^2\partial\Phi^i
\partial\Phi^i-\frac{1}{2}\lambda^2[\Phi^i,\Phi^j]^2-\left(\frac{1}{2}\lambda^2\epsilon_{ijk}\partial\Phi^i[\Phi^j,\Phi^k]\right)^2} \, \, ,
\eeq the terms under the square root can then be rewritten using the
Nahm equation as a perfect square so that\beq E=T_1\int d\sigma STr\left(I+\frac{1}{2}\lambda^2\left(\partial\Phi^i
\partial\Phi^i-\frac{1}{2}[\Phi^i,\Phi^j]^2\right)\right) \, \, ,
\eeq and we can clearly see that energy reduces to the linear form and a solution to the Nahm equation solves the full non-linear theory.

It was from this linearised starting point that
\cite{Constable:2002yn} proceeded by using the Bogomol'nyi trick to
write the energy (minus the constant piece) as 
\beq\label{linen} E=T_1\int d\sigma
STr \left( \frac{1}{2} \lambda^2 \left( \partial\Phi^i
-\frac{1}{2} c_{ijk} [\Phi^j,\Phi^k] \right)^2+T \right) \eeq 
where $T$ is
a topological piece. In order to be able to write the energy in this form one requires that\beq \label{equation}
\frac{1}{2}c_{ijk}c_{ilm}Tr\left([\Phi^j,\Phi^k][\Phi^l,\Phi^m]\right)=Tr\left([\Phi^i,\Phi^j][\Phi^i,\Phi^j]\right).
\eeq 
One can then show that the above equation (\ref{equation}) holds along
with the equations of motion derived from the action (4) if the
modified Nahm equation is obeyed along with the algebraic conditions
that follow from the vanishing of the supersymmetry variation. The
modified Nahm equation then manifestly appears as the Bogomol'nyi
equation derived by minimising (\ref{linen}).

The first configuration considered in \cite{Constable:2002yn} is with
two intersecting three-branes. This configuration requires
(\ref{ModNahm}) to be satisfied with $c_{123}=c_{145}=1$ as well as
the associated algebraic equations that follow from the
supersymmetry. Expanding the energy for five non-zero scalars we have
\beq  E=T_1\int d^2\sigma STr
\sqrt{\left(I+\frac{1}{2}\lambda^2(\partial\Phi^i
\partial\Phi^i-\frac{1}{2}[\Phi^i,\Phi^j]^2)\right)^2+\lambda^6\left(\epsilon_{ijklm}\partial\Phi^i[\Phi^j,\Phi^k][\Phi^l,\Phi^m]\right)^2}
\eeq where we have used the modified Nahm equation and associated
algebraic conditions to write the first square in that form. Thus if
the epsilon term vanishes for a solution to the linear equations of
motion, it is also solution to the full non-linear equations of
motion.  Indeed one can check that for the solutions given in
\cite{Constable:2002yn} that this is the case and so their solutions
are again  solutions of the full Born-Infeld theory. It is interesting
to note however that there do exist solutions to (\ref{ModNahm})
which do not have the form described in \cite{Constable:2002yn} where
this second term in the energy  does not vanish.  These solutions
correspond to the case where the calibration is deformed away from the
flat intersection \cite{Lambert,Helling}. The nonlinearity of the
brane action then plays a key role. In what follows we will restrict
ourselves to the case of flat intersecting branes so that the linear
equations will be enough. It would be interesting for future work to
examine the case where the calibration is deformed away from the simple intersection, from the membrane perspective.

In summary, the non-linear action is such that solutions that saturate
the bound of the linear theory are also solutions of the non-linear
theory. In the case of less than half supersymmetry there are
algebraic equations in addition to the Nahm type equation. These
algebraic equations are actually necessary to derive a Bogomol'nyi
bound. These are the guiding properties that we will use to extend the
Basu-Harvey equation and the associated membrane action to describe
more complicated five-brane configurations.

{\section{A Nahm Equation for M-theory}}

The following M-theory version of the Nahm equation was proposed  in
\cite{Basu:2004ed}, to describe membranes ending on a five-brane,
\beq\label{HarvNahm}
\pd{X^i}{s}+\frac{M_{11}^3}{64\pi}\epsilon_{ijkl}\frac{1}{4!}[G_5,X^j,X^k,X^l]=0,
\eeq where the quantum Nambu 4-bracket is defined by  
\beq
[A_1,A_2,A_3,A_4]=\sum_{permutations
\sigma}sgn(\sigma)A_{\sigma_1}A_{\sigma_2}A_{\sigma_3}A_{\sigma_4}
\eeq and $G_5$ is a difference of projection operators defined in the
Appendix. The solution to this equation as given in \cite{Basu:2004ed} is
\beq\label{HarvSol} 
X^i(s)=\frac{i\sqrt{2\pi}}{M_{11}^{3/2}}\frac{1}{\sqrt{s}}G^i\,\,
, \eeq where the set of matrices $\{ G^i \} $ are in a particular 
representation of
Spin(4) (see Appendix). Thus, the solution is
again that of a fuzzy funnel but this time there is a fuzzy three-sphere whose radius diverges to form the five-brane.

Just as the Nahm equation was generalised to describe D1 branes
ending on a three-brane that wraps a calibrated cycle so we wish to
modify the Basu-Harvey equation to describe membranes ending on a
five-brane that wraps some calibrated cycle. The ability of the
Basu-Harvey equations to be modified in a natural way so as to allow
these more general configurations will be taken as supporting evidence
in favour of the validity of their equation in describing membranes.

The natural generalisation of the Basu-Harvey equation which we propose is
\beq\label{myNahm}
\pd{X^i}{s}+\frac{M_{11}^3}{64\pi}g_{ijkl}\frac{1}{4!}[H^*,X^j,X^k,X^l]=0,
\eeq where $g_{ijkl}$ is an anti-symmetric constant tensor that is
associated to the components of the relevant calibration form that
represents an intersecting five-brane configuration. $H^*$ has
properties analogous to those of $G_5$, namely $(H^*)^2=1$ and for the
solutions we consider $\{H^*,X^i\}=0$.

In \cite{Basu:2004ed} an expression for the membrane energy in such a
configuration was also postulated,  \beq E=T_2\int
d^2\sigma Tr\left[\left( X^{i^
\prime}+\epsilon_{ijkl}\frac{1}{4!}[G_5,X^j,X^k,X^l]\right)^2+\left(1-\frac{1}{2}\epsilon_{ijkl}\{{X^{i^\prime}},\frac{1}{4!}[G_5,X^j,X^k,X^l]\}\right)^2\right]^{1/2}.
\eeq  Now the $G_5$'s drop out of this expression, and using the
Basu-Harvey equation it can be rewritten as \beq E=T_2\int
d^2\sigma Tr\left[\left(1+\frac{1}{2}(\partial_a
X^i)^2-\frac{1}{2.3!}[X^i,X^j,X^k]^2\right)^2\right]^{1/2} \eeq  where
$[X^i,X^j,X^k]$ is a Nambu 3-bracket containing all permutations of
the three entries with signs. So for three active scalars if the
Basu-Harvey equation is obeyed the action proposed in
\cite{Basu:2004ed} is equivalent to its {\it{linearised}} version
\beq\label{3action} S=-T_2\int d^3\sigma
Tr\left(1+\frac{1}{2}(\partial_a
X^i)^2-\frac{1}{2.3!}[X^j,X^k,X^l]^2\right) \, \, . \eeq

We propose that when more scalars are
activated we can use our modified
Basu-Harvey equation (\ref{myNahm}) to rewrite the action as the
linear piece squared plus other squared terms. When these other terms
are zero, the {\it{linearised}} action is the full action. The energy is
then given by \beq E=\frac{T_2}{2}\int d^2\sigma Tr\left( X^{i^ \prime}
X^{i^\prime }-\frac{1}{3!}[X^j,X^k,X^l][X^j,X^k,X^l]\right) \eeq where
we have subtracted the constant piece and assumed that the $X^i$
depend only on $\sigma_2$($=X^{10}$ in this gauge). Following the
usual Bogomol'nyi construction we rewrite this as \beq\label{energy}
E=\frac{T_2}{2}\int d^2\sigma \left\{Tr\left(
X^{i^\prime}+g_{ijkl}\frac{1}{4!}[H^*,X^j,X^k,X^l]\right)^2+T\right\}
\eeq where $T$ is a {\it{topological}} piece given by \beq T=-T_2\int
d^2\sigma
Tr\left(\frac{M_{11}^3}{64\pi}g_{ijkl}\pd{X^i}{\sigma_2}\frac{1}{4!}[H^*,X^j,X^k,X^l]\right)
\eeq (with factors restored). This reproduces the correct energy
density for the five-brane on which the membranes end in the case where
$g_{ijkl}=\epsilon_{ijkl}$. Now, to rewrite in the energy in this way
we have imposed that \bea
&&\frac{1}{3!}g_{ijkl}g_{ipqr}Tr\left([H^*,X^j,X^k,X^l][H^*,X^p,X^q,X^r]\right)\\
\nonumber &&=Tr\left([H^*,X^i,X^j,X^k][H^*,X^i,X^j,X^k]\right) \eea
which using $\{H^*,X^i\}=0$ and $(H^*)^2=1$ is equivalent to the
following algebraic constraint 
\beq\label{constraint}
\frac{1}{3!}g_{ijkl}g_{ipqr}Tr\left([X^j,X^k,X^l][X^p,X^q,X^r]\right)=Tr\left([X^i,X^j,X^k][X^i,X^j,X^k]\right)
\, \, .  
\eeq 
Note, this is satisfied for the case
$g_{ijkl}=\epsilon_{ijkl}$ when the only scalars activated are $X^2$
to $X^5$. The Bogomol'nyi equation found from minimising the energy
given in equation (\ref{energy}) is then our modified
Basu-Harvey equation (\ref{myNahm}).

We now show explicitly that the equation of motion following from the
action  (\ref{3action}) when combined with the modified Basu-Harvey
equation imply  the constraint (\ref{constraint}).

The equation of motion is given by \beq
X^{i^{\prime\prime}}=-\frac{1}{2}\lt X^j,X^k,[X^i,X^j,X^k]\rt \eeq
where the three bracket $\lt A,B,C \rt$ is the sum of the six
permutations of the three entries, but with the sign of the permutation
determined only by the order of the first two entries, i.e. $ABC, ACB
\mbox{ and } CAB$ are the positive permutations.  By using the
Bogomol'nyi equation (\ref{myNahm}) twice on the left-hand side  we
get: 
\beq
\label{3bconst} \frac{1}{3!} g_{ijkl} g_{jpqr}\lt
X^k,X^l,[X^p,X^q,X^r]\rt=-\lt X^j,X^k,[X^i,X^j,X^k]\rt \, \, .  
\eeq
After multiplying by $X^i$ and taking the trace, we recover the above
constraint equation, (\ref{constraint}). Thus in summary, the
solutions of the generalised Basu-Harvey equation (\ref{myNahm}) that
obey the constraint equation (\ref{3bconst}) are solutions to the
equations of motion of the proposed membrane action (\ref{3action}).

\medskip

\section{Supersymmetry}

In the D-brane case, both the Nahm like equation and the additional
algebraic relations could be derived from imposing that the necessary
supersymmetry variation vanished. The approach we have described above
is equivalent but less efficient. The Bogomol'nyi argument
effectively implies the Nahm type equation and the necessary
algebraic relations. We would like to encode this information by
imposing by fiat a supersymmetry transformation whose vanishing will
imply the above equations. As to whether this can be made more
concrete by constructing a supersymmetric membrane action with this
supersymmetry variation we leave as an interesting open question. The
advantage of this imposed supersymmetry variation is that it will
allow us to relate the solutions of the membrane equations to
intersecting five-branes that preserve various fractions of
supersymmetry.
The obvious suggested susy variation is 
\beq  \delta\lambda=\left(\frac{1}{2}\partial_\mu
X^i\Gamma^{\mu
i}-\frac{1}{2.4!}\gtb{i}{j}{k}\Gamma^{ijk}\right)\epsilon.  \eeq 

Now, we use the modified Basu-Harvey equation in the first term and
rearrange so that the requirement that the supersymmetry variation
vanishes becomes that
\beq\label{susycond}
\sum_{i<j<k}\tb{i}{j}{k}\Gamma^{ijk}(1-g_{ijkl}\Gamma^{ijkl\#})\epsilon=0,
\eeq
we have removed and overall factor of $H*$ from the left-hand side
since, like $G_5$, it is the difference of projection operators onto
orthogonal sub-spaces and has trivial kernel. $\epsilon$ is the preserved supersymmetry on the membrane \wv and we
have $\Gamma^{01\#}\epsilon=\epsilon$ where the membrane's worldvolume
is in the 0, 1 and $10=\#$ directions. We can then solve the
supersymmetry condition (\ref{susycond}) by defining projectors
\beq
\label{proj} P_{ijkl}=\frac{1}{2}(1-g_{ijkl}\Gamma^{ijkl\#}) 
\eeq
where there is no sum over $i,j,k \mbox{ or }l$. We normalise
$g_{ijkl}=\pm1$ so they obey $P_{ijkl}P_{ijkl}=P_{ijkl}$. (Note, in all the
cases that we will consider, for each triplet $i,j,k$, $g_{ijkl}$ is only
non-zero for at most one value of $l$). We impose $P_{ijkl}\epsilon=0$
for each $i,j,k,l$ such that $g_{ijkl}\neq0$. Then by using
the membrane projection ($\Gamma^{01\#}\epsilon=\epsilon$) we can see
that each projector $P_{ijkl}$ corresponds to a five-brane in the
$0,1,i,j,k,l$ directions. To apply the projectors simultaneously, the matrices $\Gamma_{ijkl\#}$ need to commute with each other. $[\Gamma_{ijkl\#},\Gamma_{i'j'k'l'\#}]=0$ if and only if the
sets $\{i,j,k,l\}$ and  $\{i',j',k',l'\}$ have two or zero elements in
common, corresponding to five-branes intersecting over a three-brane
soliton or a string soliton.

Once we impose the set of mutually commuting projectors, the
supersymmetry transformation (\ref{susycond}) reduces to
\beq\label{susyids} \sum_{g_{ijkl}=
0}\tb{i}{j}{k}\Gamma^{ijk}\epsilon=0.  
\eeq
Here we sum over triplets $i,j,k,$ such that $g_{ijkl}=0$ for
all $l$. Using the projectors allows us to express these as a set of
conditions on the 3-brackets alone.

\section{Five-Brane Configurations}

We will now describe the specific equations that emerge which correspond 
to the various possible intersecting five-brane configurations. 

The five-branes must always have at least one spatial direction in common 
corresponding to the direction in which the membrane intersects the five-branes. These configurations of five-branes can also be thought
of as a single five-brane stretched over a calibrated manifold \cite{Harvey:1982xk}. These
five-brane intersections can be found in 
\cite{Gibbons:1998hm,Jerome,Acharya:1998en}. We list the
conditions following from the modified Basu-Harvey Equation, those following 
from the supersymmetry conditions
(\ref{susyids}) (with $\nu$ the fraction of preserved supersymmetry) and then 
discuss any remaining conditions required to
satisfy the constraint (\ref{3bconst}). In string case, \cite{Constable:2002yn}
only the supersymmetry conditions and the Jacobi identity were
required to satisfy the equivalent constraint.

\subsection{Configuration 1}

The first configuration corresponding to the single five-brane \cite{Basu:2004ed} is

\bea\label{c1} \matrix{ M5:&1&2&3&4&5\cr
M2:&1&&&&&&&&&&\#\cr}\nonumber\\ g_{2345}=1\quad\quad\quad
\nu=1/2\quad {} \eea \bea {X^2}' = -H^*\tb{3}{4}{5}&,& \quad {X^3}' =
H^*\tb{4}{5}{2}\ , \quad \nonumber\\ {X^4}' = -H^*\tb{5}{2}{3}&,&
\quad {X^5}' = H^*\tb{2}{3}{4}\ . \nonumber\\ \nonumber \eea

\subsection{Configuration 2}

For the next case we consider two five-branes intersecting on a three-brane corresponding to an SU(2) Kahler calibration of a two surface embedded in four dimensions. The activated scalars are $X^2$ to $X^7$.

\bea\label{c2} \matrix{ M5:&1&2&3&4&5\cr M5:&1&2&3&&&6&7\cr
M2:&1&&&&&&&&&&&\#\cr}\nonumber\\
g_{2345}=g_{2367}=1\quad\quad\quad\quad\qquad \nu=1/4\quad {} \eea
\bea {X^2}' =-H^*\tb{3}{4}{5} -H^*\tb{3}{6}{7} &,& \quad {X^3}' =
H^*\tb{4}{5}{2} +H^*\tb{6}{7}{2}\ , \quad\nonumber\\ {X^4}' =
-H^*\tb{5}{2}{3} &,& \quad {X^5}' = H^*\tb{2}{3}{4}\ , \quad
\nonumber\\ {X^6}' = -H^*\tb{7}{2}{3} &,& \quad {X^7}' =
H^*\tb{2}{3}{6}\ , \nonumber\\ \tb{2}{4}{6}= \tb{2}{5}{7} &,&  \quad
\tb{2}{5}{6}=-\tb{2}{4}{7}\ , \quad \nonumber\\ \tb{3}{4}{6}=
\tb{3}{5}{7} &,&  \quad \tb{3}{5}{6}=-\tb{3}{4}{7}, \quad \nonumber\\
\tb{4}{5}{6}=\tb{4}{5}{7}&=&\tb{4}{6}{7}=\tb{5}{6}{7}=0. \quad
\nonumber \eea

In order to satisfy the constraint we need the $X^i$'s to satisfy the
following equations:

$$ \mbox{Choose } m\in\{2,3\},\quad i,j,k,l\in\{4,5,6,7\}\nonumber\\
$$\bea \epsilon_{ijk}\fbr{i}{m}{m}{j}{k}&=&0,\quad(\mbox{no sum over }
m)\nonumber\\ \epsilon_{ijkl}\fbr{i}{j}{m}{k}{l}&=&0.\label{jac1} \eea

In the string theory case there were no additional equations, as
apart from the Nahm like equations and algebraic conditions on the
brackets all that was needed to solve the constraint was the Jacobi
identity. If $X^m$ anti-commutes with $X^i,X^j,X^k$ then
the first equation reduces to the Jacobi identity. Similarly if $X^m$
anti-commutes with $X^i,X^j,X^k,X^l$ the second equation reduces to
\beq
 \epsilon_{ijkl}X^iX^jX^kX^l=0.
\eeq
\subsection{Configuration 3}

Three five-branes can intersect on a three-brane corresponding to an SU(3) Kahler
calibration of a two surface embedded in six dimensions. The active scalars are $X^2$ to $X^9$.

\bea\label{c3} \matrix{ M5:&1&2&3&4&5\cr M5:&1&2&3&&&6&7\cr
M5:&1&2&3&&&&&8&9\cr M2:&1&&&&&&&&&\#\cr}\nonumber\\
g_{2345}=g_{2367}=g_{2389}=1\quad\quad\quad\qquad \nu=1/8\quad {} \eea
\bea {X^2}' =-H^*\tb{3}{4}{5} &-&H^*\tb{3}{6}{7}-H^*\tb{3}{8}{9} ,
\quad\nonumber\\ {X^3}' = H^*\tb{4}{5}{2}
&+&H^*\tb{6}{7}{2}+H^*\tb{8}{9}{2}\ , \quad\nonumber\\ {X^4}' =
-H^*\tb{5}{2}{3} &,& \quad {X^5}' = H^*\tb{2}{3}{4}\ , \quad
\nonumber\\ {X^6}' = -H^*\tb{7}{2}{3} &,& \quad {X^7}' =
H^*\tb{2}{3}{6}\ , \nonumber\\ {X^8}' = -H^*\tb{9}{2}{3} &,& \quad
{X^9}' = H^*\tb{2}{3}{8}\ , \nonumber\\ \tb{2}{4}{6}= \tb{2}{5}{7} &,&
\quad \tb{2}{5}{6}=-\tb{2}{4}{7}, \quad \nonumber\\ \tb{2}{4}{8}=
\tb{2}{5}{9} &,&  \quad  \tb{2}{5}{8}=-\tb{2}{4}{9}\ , \quad
\nonumber\\ \tb{2}{6}{8}= \tb{2}{7}{9} &,&  \quad
\tb{2}{7}{8}=-\tb{2}{6}{9}, \quad \nonumber\\ \tb{3}{4}{6}=
\tb{3}{5}{7} &,& \quad \tb{3}{5}{6}=-\tb{3}{4}{7}, \quad \nonumber\\
\tb{3}{4}{8}= \tb{3}{5}{9} &,&  \quad  \tb{3}{5}{8}=-\tb{3}{4}{9}\ ,
\quad \nonumber\\ \tb{3}{6}{8}= \tb{3}{7}{9} &,&  \quad
\tb{3}{7}{8}=-\tb{3}{6}{9}, \quad \nonumber\\
\tb{4}{5}{6}+\tb{6}{8}{9}=0\ &,&\quad \tb{4}{5}{7}+\tb{7}{8}{9}=0\
,\quad \nonumber\\ \tb{4}{5}{8}+\tb{6}{7}{8}=0\ &,&\quad
\tb{4}{5}{9}+\tb{6}{7}{9}=0\ ,\quad \nonumber\\
\tb{4}{6}{7}+\tb{4}{8}{9}=0\ &,&\quad \tb{5}{6}{7}+\tb{5}{8}{9}=0\
,\quad \nonumber\\
\tb{4}{6}{8}=\tb{4}{7}{9}&+&\tb{5}{6}{9}+\tb{5}{7}{8}\ ,
\quad\nonumber\\
\tb{5}{7}{9}=\tb{5}{6}{8}&+&\tb{4}{7}{8}+\tb{4}{6}{9}\ .
\quad\nonumber \eea

In order to satisfy the constraint again we have additional algebraic
constraints for certain $X^i$'s, for this we define ``pairs'' as
$\{2,3\}$,$\{4,5\}$, $\{6,7\}$ and $\{8,9\}$
$$ \mbox{Choose } m\in\{2,3\},\quad i,j,k,l\in\{4,5,6,7,8,9\}\mbox{
such that } \{i,j\},\{k,l\} \mbox{ are pairs}\nonumber\\
$$\bea \epsilon_{ijk}\fbr{i}{m}{m}{j}{k}&=&0,\quad(\mbox{no sum over }
m)\nonumber\\ \epsilon_{ijkl}\fbr{i}{j}{m}{k}{l}&=&0. \label{jac3} \eea

\subsection{Configuration 4}

The next configuration has 3 five-branes intersecting over a string which corresponds to an SU(3) Kahler calibration of a four surface in six dimensions.
There are only 6 activated scalars.

\bea\label{c4} \matrix{ M5:&1&2&3&4&5\cr M5:&1&2&3&&&6&7\cr
M5:&1&&&4&5&6&7\cr M2:&1&&&&&&&&&\#\cr}\nonumber\\
g_{2345}=g_{2367}=g_{4567}=1\quad\quad\quad\qquad \nu=1/8\quad {} \eea
\bea {X^2}' &=&-H^*\tb{3}{4}{5} -H^*\tb{3}{6}{7}\ , \quad\nonumber\\
{X^3}' &=& H^*\tb{2}{4}{5} +H^*\tb{2}{6}{7}\ , \quad\nonumber\\ {X^4}'
&=&-H^*\tb{2}{3}{5} -H^*\tb{5}{6}{7}\ , \quad\nonumber\\ {X^5}' &=&
H^*\tb{2}{3}{4} +H^*\tb{4}{6}{7}\ , \quad\nonumber\\ {X^6}'
&=&-H^*\tb{2}{3}{7} -H^*\tb{4}{5}{7}\ , \quad\nonumber\\ {X^7}' &=&
H^*\tb{2}{3}{6} +H^*\tb{4}{5}{6}\ , \quad\nonumber\\
\tb{2}{4}{6}&=&\tb{2}{5}{7}+\tb{3}{5}{6}+\tb{3}{4}{7}\ ,
\quad\nonumber\\
\tb{3}{5}{7}&=&\tb{3}{4}{6}+\tb{2}{5}{6}+\tb{2}{4}{7}\ .
\quad\nonumber \eea

In order to satisfy the constraint again we have to satisfy additional
algebraic constraints for certain $X^i$'s, for this we define
``pairs'' as $\{2,3\}$,$\{4,5\}$ and $\{6,7\}$
$$ \mbox{Choose } i,j,k,l,m\in\{2,3,4,5,6,7\}\mbox{ such that }
\{i,j\},\{k,l\} \mbox{ are pairs}\nonumber\\
$$\bea \epsilon_{ijk}\fbr{i}{m}{m}{j}{k}&=&0,\quad(\mbox{no sum over }
m)\nonumber\\ \epsilon_{ijkl}\fbr{i}{j}{m}{k}{l}&=&0.\label{jac4}  \eea

\subsection{Configuration 5}

In the next configuration we are forced by supersymmetry to have an
additional anti-brane. Even though there are only three independent projectors this configuration has three five-branes and an anti-five-brane intersecting over a membrane. This corresponds to the SU(3) special Lagrangian calibration of a three surface embedded in six dimensions.

\bea\label{c5} \matrix{ M5:&1&2&3&4&5\cr M5:&1&2&&4&&6&&8\cr
\bar{M5}:&1&2&3&&&6&7\cr M5:&1&2&&&5&&7&8\cr
M2:&1&&&&&&&&&\#\cr}\nonumber\\
g_{2345}=g_{2468}=-g_{2367}=g_{2578}=1\quad\quad\qquad \nu=1/8\quad {}
\eea \bea {X^2}' =-H^*\tb{3}{4}{5} &-&H^*\tb{4}{6}{8} +H^*\tb{3}{6}{7}
-H^*\tb{5}{7}{8}\ , \nonumber\\ {X^3}' &=& H^*\tb{2}{4}{5}
-H^*\tb{2}{6}{7}\ , \quad\nonumber\\ {X^4}' &=&-H^*\tb{2}{3}{5}
-H^*\tb{2}{6}{8}\ , \quad\nonumber\\ {X^5}' &=& H^*\tb{2}{3}{4}
+H^*\tb{2}{7}{8}\ , \quad\nonumber\\ {X^6}' &=& H^*\tb{2}{3}{7}
-H^*\tb{2}{4}{8}\ , \quad\nonumber\\ {X^7}' &=&-H^*\tb{2}{3}{6}
-H^*\tb{2}{5}{8}\ , \quad\nonumber\\ {X^8}' &=& H^*\tb{2}{4}{6}
+H^*\tb{2}{5}{7}\ , \quad\nonumber\\ \tb{3}{4}{7}&=&\tb{3}{5}{6}\
,\quad \tb{4}{3}{8}=\tb{4}{5}{6}\ ,\quad\nonumber\\
\tb{5}{3}{8}&=&\tb{5}{7}{4}\ ,\quad \tb{6}{4}{7}=\tb{6}{8}{3}\
,\quad\nonumber\\ \tb{7}{3}{8}&=&\tb{7}{6}{5}\ ,\quad
\tb{8}{4}{7}=\tb{8}{6}{5}\ ,\quad\nonumber\\
\tb{2}{3}{8}&+&\tb{2}{4}{7}+\tb{2}{6}{5}=0\ , \quad\nonumber\\
\tb{6}{7}{8}&+&\tb{4}{5}{8}+\tb{3}{4}{6}+\tb{3}{5}{7}=0\ .
\quad\nonumber \eea

In order to satisfy the constraint once again we have similar
additional algebraic constraints for certain $X^i$'s, for this we
define ``pairs'' as $\{3,8\}$,$\{4,7\}$ and $\{5,6\}$
$$ \mbox{Choose }m\in\{2,\dots,8\},\ i,j,k,l,\in\{3,4,5,6,7,8\}\mbox{
such that } \{i,j\},\{k,l\} \mbox{ are pairs}\nonumber
$$ \bea \epsilon_{ijk}\fbr{i}{m}{m}{j}{k}&=&0,\quad(\mbox{no sum over
} m)\nonumber\\ \epsilon_{ijkl}\fbr{i}{j}{m}{k}{l}&=&0.\label{jac5}
\eea

\section{Solutions}

The Basu-Harvey equation is solved by
\beq 
X^i(s)=\frac{i\sqrt{2\pi}}{M_{11}^{3/2}}
\left(\frac{2n+6}{2n+1}\right)^{1/2}\frac{1}{\sqrt{s}}G^i 
\eeq
where $n$ labels the specific representation of $Spin(4)$ (see the Appendix). We can solve the cases of intersecting five-branes analogously to the intersecting three-branes of \cite{Constable:2002yn} by effectively using multiple copies of this solution. The first multi-five-brane case (\ref{c2}) is solved by setting
\beq 
X^i(s)=\frac{i\sqrt{2\pi}}{M_{11}^{3/2}}\left(\frac{2n+6}{2n+1}\right)^{1/2}\frac{1}{\sqrt{s}}H^i 
\eeq
where the $H^i$ are given by the block-diagonal $2N\times 2N$ matrices
\bea H^2 &=& \mbox{diag}\, (G^1,G^1)\nonumber\\ H^3 &=& \mbox{diag}\,
(G^2,G^2)\nonumber\\ H^4 &=& \mbox{diag}\, (G^3,0)\nonumber\\ H^5 &=&
\mbox{diag}\, (G^4,0)\nonumber\\ H^6 &=& \mbox{diag}\,
(0,G^3)\nonumber\\ H^7 &=& \mbox{diag}\, (0,G^4)\nonumber\\ H^* &=&
\mbox{diag}\, (G^5,G^5)\, , 
\eea
which are such that
\beq 
H^i+\frac{n+3}{8(2n+1)}g_{ijkl}\frac{1}{4!}[H^*,H^j,H^k,H^l]=0.  
\eeq
This makes sure that the conditions following
from the generalised Basu-Harvey equation vanish. The remaining
conditions in (\ref{c2}), that is those following from the
supersymmetry transformation, are satisfied trivially as the three
brackets involved all vanish for this solution. The first additional algebraic equation of (\ref{jac1}) is satisfied for the solution as the indices must be chosen such that for each diagonal block at least one of the $X^i$'s appearing in the bracket that has a zero there, thus the term with each permutation
vanishes independently. Again the second additional algebraic equation is trivially satisfied as there are no non-zero products of $5$ different $X^i$'s.

The more complicated cases follow easily: configuration 3 is given by
the block-diagonal $3N\times 3N$ matrices

\bea H^2 &=& \mbox{diag}\, (G^1,G^1,G^1)\nonumber\\ H^3 &=&
\mbox{diag}\, (G^2,G^2,G^2)\nonumber\\ H^4 &=& \mbox{diag}\,
(G^3,0,0)\nonumber\\ H^5 &=& \mbox{diag}\, (G^4,0,0)\nonumber\\ H^6
&=& \mbox{diag}\, (0,G^3,0)\nonumber\\ H^7 &=& \mbox{diag}\,
(0,G^4,0)\nonumber\\ H^8 &=& \mbox{diag}\, (0,0,G^3)\nonumber\\ H^9
&=& \mbox{diag}\, (0,0,G^4)\nonumber\\ H^* &=& \mbox{diag}\,
(G^5,G^5,G^5)\, , \eea

and configuration 4 by

\bea H^2 &=& \mbox{diag}\, (G^1,G^1,0)\nonumber\\ H^3 &=&
\mbox{diag}\, (G^2,G^2,0)\nonumber\\ H^4 &=& \mbox{diag}\,
(G^3,0,G^1)\nonumber\\ H^5 &=& \mbox{diag}\, (G^4,0,G^2)\nonumber\\
H^6 &=& \mbox{diag}\, (0,G^3,G^3)\nonumber\\ H^7 &=& \mbox{diag}\,
(0,G^4,G^4)\nonumber\\ H^* &=& \mbox{diag}\, (G^5,G^5,G^5)\, . \eea

Configuration 5 is

\bea H^2 &=& \mbox{diag}\, (G^1,G^1,G^1,G^1)\nonumber\\ H^3 &=&
\mbox{diag}\, (G^2,0,G^2,0)\nonumber\\ H^4 &=& \mbox{diag}\,
(G^3,G^2,0,0)\nonumber\\ H^5 &=& \mbox{diag}\,
(G^4,0,0,G^2)\nonumber\\ H^6 &=& \mbox{diag}\,
(0,G^3,G^4,0)\nonumber\\ H^7 &=& \mbox{diag}\,
(0,0,G^3,G^3)\nonumber\\ H^8 &=& \mbox{diag}\,
(0,G^4,0,G^4)\nonumber\\ H^* &=& \mbox{diag}\, (G^5,G^5,G^5,G^5) \, . \eea

\section*{Acknowledgements}
We wish to thank Anirban Basu, Jeff Harvey, Sanjaye Ramgoolam and
David Tong for discussions. DSB is supported by EPSRC grant
GR/R75373/02 and would like to thank DAMTP and Clare Hall college
Cambridge for continued support. NBC is supported by a PPARC
studentship. This work was in part supported by the EC Marie Curie
Research Training Network, MRTN-CT-2004-512194.

\appendix
\section{The Fuzzy 3-Sphere}

To make this self-contained we include a brief description of the
fuzzy 3-sphere construction following that in \cite{Basu:2004ed}. The
approach was developed in
\cite{Guralnik:2000pb,Ramgoolam:2001zx,Ramgoolam:2002wb,Sheikh-Jabbari:2004ik},
with an interesting string interpretation further developed in \cite{SJ2}.

Unlike even fuzzy spheres we must use reducible representations of
$spin(4) =SU(2) \times SU(2)$.  $\cRp$ and $\cRm$ are the
$(\frac{n+1}{4},\frac{n-1}{4})$ and $(\frac{n-1}{4},\frac{n+1}{4})$
representations respectively where $n$ is an odd integer. The
dimension of $\cR = \cRp \oplus \cRm$ is given by $N=(n+1)(n+3)/2$.

The coordinates on the fuzzy $S^3$ are  the $N  \times N$ matrices
$G^i$ ($i=1$ to 4). These matrices are defined by \beq G^i =  \Prp
\sum_{s=1}^n \rho_s (\Gamma^i P_-) \Prm + \Prm  \sum_{s=1}^n \rho_s
(\Gamma^i P_+) \Prp,  \eeq    where \beq \sum_{s=1}^n \rho_s
(\Gamma^i) = (\Gamma^i \otimes \ldots \otimes 1 +\ldots +1 \otimes
\ldots \otimes \Gamma^i )_{\rm{sym}},  \eeq where sym stands for the
completely symmetrised $n-$fold tensor product  representation of
$spin(4)$. Here $P_{\pm} = \frac{1}{2} (1 \pm \Gamma_5)$, and $\Prp,
\Prm$ are projection operators onto the irreducible  representations
$\cRp, \cRm$ respectively of $spin(4)$. The matrix $G_5$ which is
important to the construction is given by \beq G_5 = \Prp - \Prm.
\eeq (Some intuition can be gained from the fact that for $n=1$, the
matrices $G^i$ and $G_5$ become $\Gamma^i$ and $\Gamma_5$
respectively.)

The $G^i$ are elements of $End(\cR)$. We can write $G^i = G^i_+ +
G^i_-$ with $G^i_\pm = \frac{1}{2} (1 \pm G_5) G^i$ and then $G^i_\pm$
act as homomorphisms from $\cR_\mp$ to $\cR_\pm$.

From the above definitions, after much manipulation \cite{Basu:2004ed}
we can obtain the equation \beq G^i +\frac{n+3}{8(2n+1)}
\epsilon_{ijkl} G_5 G^j G^k G^l =0.  \eeq Thus the Basu-Harvey
equation (\ref{HarvNahm}) is solved by (\ref{HarvSol}) in
the large $N$ limit. However a solution can be found for any $n$ by
taking
\beq X^i(s)=\frac{i\sqrt{2\pi}}{M_{11}^{3/2}}\left(\frac{2n+6}{2n+1}\right)^{1/2}\frac{1}{\sqrt{s}}G^i\,\,
. \eeq 




\end{document}